\documentclass[11pt]{article}
\usepackage{amsmath}
\usepackage{amssymb}
\usepackage{amsfonts}
\usepackage{amsthm}
\usepackage{graphicx} 
\usepackage{subfigure}
\usepackage{enumerate}
\usepackage{verbatim}
\usepackage{authblk}
\usepackage{setspace}
\usepackage{enumitem}
\begin{document}
%\linenumbers
\title{Combinatorial Batch Codes: A Lower Bound and Optimal Constructions}
\author[1]{Srimanta Bhattacharya}
\author[2]{Sushmita Ruj}
\author[1]{Bimal Roy}
\affil[1]{Applied Statistics Unit, Indian Statistical Institute,\authorcr
203 B T Road, Kolkata 700 108. India\authorcr
Email: \{srimanta\_r, bimal\}@isical.ac.in
}
\affil[2]{School of Information Technology and Engineering, \authorcr
University of Ottawa,\authorcr
Ottawa, K1N6N5. Canada\authorcr
Email: sruj@site.uottawa.ca
}
\date{}
\maketitle
\newcommand{\pf}{\noindent{\em Proof. }}
\newcommand{\cn}{\noindent{\em {\bf Construction. }}}
\newcommand{\pocr}{\noindent{\em Proof of correctness. }}
\newcommand{\an}{\noindent{\em Analysis. }}
\newtheorem{theorem}{\bf Theorem}[section]
\newtheorem{proposition}{\bf Proposition}[section]
\newtheorem{lemma}{\bf Lemma}[section]
\newtheorem{corollary}{\bf Corollary}[section]
\newtheorem{definition}{\bf Definition}[section]
\theoremstyle{definition}
\newtheorem{example}{\bf Example}[subsection]
\theoremstyle{remark}
\newtheorem{remark}{Remark}[subsection]
\newtheoremstyle{simple}
{5pt}
{5pt}
{\itshape}
{}
{\itshape}
{}
{1pt}
{}
\theoremstyle{simple}
\newtheorem{brtheorem}{\bf Theorem}[section]
\newtheorem*{nhtheorem}{}

\newcommand{\comb}[2] {\mbox{$\left( { #1 \atop #2 } \right)$}}
\newcommand{\card}[1]{\mbox{$\mid #1 \mid$}}
\newcommand{\ovx}{\overline x}
\newcommand{\ovv}{\overline v}
\newcommand{\ovw}{\overline w}
\newcommand{\ovu}{\overline u}
\newcommand{\ovf}{\overline f}
\renewcommand{\thefootnote}{\fnsymbol{footnote}}
\numberwithin{equation}{section}
%\tableofcontents
\begin{abstract}
Batch codes, introduced by Ishai, Kushilevitz, Ostrovsky and Sahai in ~\cite{YuKuOsSa}, are methods for solving the following data storage problem: $n$ data items are to be stored in $m$ servers in such a way that any $k$ of the $n$ items can be retrieved by reading at most $t$ items from each server, and that the total number of items stored in $m$ servers is $N$. A {\em Combinatorial batch code} (CBC) is a batch code where each data item is stored without change, i.e., each stored data item is a copy of one of the $n$ data items.\par
One of the basic yet challenging problems is to find {\em optimal} CBCs, i.e., CBCs for which total storage ($N$) is minimal for given values of $n$, $m$, $k$, and $t$. In \cite{PaStWe}, Paterson, Stinson and Wei exclusively studied CBCs and gave constructions of some optimal CBCs. \par
 In this article, we give a lower bound on the total storage ($N$) for CBCs. We give explicit construction of optimal CBCs for a range of values of $n$. For a different range of values of $n$, we give explicit construction of optimal and almost optimal CBCs.  Our results partly settle an open problem of \cite{PaStWe}.
\end{abstract}

\section{Introduction}
Batch codes, introduced  by Ishai, Kushilevitz, Ostrovsky and Sahai in \cite{YuKuOsSa}, concerns the problem of distributing a database of $n$ items among $m$ servers in such a way that any $k$ of the $n$ items can be retrieved by reading at most $t$ items from each server, while keeping the total storage over $m$ servers to $N$. In \cite{YuKuOsSa}, the authors formalized the notion of a batch code with the following general definition.

\begin{definition}{\bfseries ~\cite{YuKuOsSa}}
An $(n, N, k, m, t)$ batch code over an alphabet $\Sigma$ is defined by an encoding function $C : {\Sigma}^n \rightarrow ({\Sigma}^{*})^{m}$ (each output of which is called a bucket) and a decoding algorithm $A$ such that:
\begin{enumerate}
\item The total length of all $m$ buckets is $N$ (where the length of each bucket is independent of $x \in {\Sigma}^n$);
\item For any $x \in {\Sigma}^n$ and $\{i_1 , \ldots, i_k\} \subseteqq [n], A(C(x), i_1 , \ldots, i_k) = (x_{i_1} , \ldots, x_{i_k})$, and $A$ probes at most $t$ symbols from each bucket in $C(x)$ (whose positions are determined by $ i_1, \ldots, i_k $ ).
\end {enumerate}
\end{definition}
In the definition, a string of length $n$ (i.e., ``$x \in {\Sigma}^n$'') corresponds to a set of $n$ data items, ``$m$ {\em buckets}'' refers to $m$ sets of data items stored in $m$ servers and ``total length'' refers to total storage ($N$). As part of an encoding algorithm, one can apply any suitable transformation on the data items (e.g., $XOR$ for binary data) to be stored, with the condition that the corresponding decoding algorithm should make it possible to retrieve any subset (of prespecified size ($k$)) of data items by reading a limited ($t$) number of items from each server. Apart from the defining parameters ($n, N, k, m, t$), another important parameter is {\em rate}. Rate of a batch code is defined as the ratio $\frac{n}{N}$. \par
Since a fixed number of items are read from each server while retrieving a batch of items, batch codes can be used for balancing load among servers in a distributed database scenario. It was shown in \cite{YuKuOsSa} that batch codes can also be used for amortizing computational overhead in private information retrieval protocols.\par
Of particular interest is the class of batch codes for which encoding is assignment (storage) of items to servers and decoding is retrieval (reading) of items from servers. This class of batch codes are called {\em replication-based batch codes} \cite{YuKuOsSa} or {\em combinatorial batch codes (CBC)} \cite{PaStWe}. One obvious advantage of CBCs is that their encoding and decoding do not incur additional computational overhead. On the other hand, storage requirement may be more for CBCs. This fact is illustrated by a nice example in the introduction of \cite{YuKuOsSa}. As combinatorial objects CBCs are quite interesting, and so far they have been studied in various combinatorial frameworks. In \cite{YuKuOsSa}, Ishai et al. used the framework of unbalanced expanders. In \cite{PaStWe}, Paterson et al. studied CBCs in the setting of set systems. In \cite{BrKiMeSc}, Brualdi et al. explored the connection between CBCs and transversal matroids.\par
Before proceeding further, we mention here that in this article, we will exclusively consider CBCs with $t=1$ and will not explicitly include this parameter in any expression. For example, we will write $(n, N, k, m)$-CBC to denote an $(n, N, k, m, t)$-CBC with $t=1$. \par
An $(n, N, k, m)$-CBC is called {\em optimal} if $N$ is minimal for given $n, m,$ and $k$. We denote by $N(n, m, k)$ value of $N$ for an optimal $(n, N, k, m)$-CBC. So, for any $(n, N', k, m)$-CBC it follows that $N' \geq N(n, k, m)$. An interesting and practically important problem is to find optimal CBCs: given $n, m, k$, the objective is to find $N(n, m, k)$ and to give explicit construction of an optimal $(n, N(n, k, m), k, m)$-CBC. For example, if $n \geq m$, then it may be trivially observed that $N(n, k, m) = n$, and for the corresponding optimal CBC, $n$ items are stored in any $n$ out of $m$ servers. But for $n \geq m+1$, finding optimal CBCs is a fairly non-trivial problem. In \cite{PaStWe} and \cite{BrKiMeSc}, this problem was addressed and some partial results were obtained. Next, we briefly discuss these results.\par
\begin{brtheorem}[{\bf \cite{PaStWe}}]
$N(n, k, k) = kn-k(k-1)$.
\label{thm1}
\end{brtheorem}
For optimal CBC with the above parameters, items are stored in the following way:
\begin{enumerate}[topsep=0pt, label=(\roman*)]
\setlength{\itemsep}{0pt}
\setlength{\parskip}{0pt}
\setlength{\parsep}{0pt}
\setlength{\topsep}{0pt}
\setlength{\partopsep}{0pt}
\setlength{\topskip}{0pt}
\setlength{\headsep}{0pt}
\setlength{\topmargin}{0pt}
\setlength{\itemindent}{7pt}
\setlength{\listparindent}{7pt}
\item Any $k$ of the $n$ items are stored in $k$ servers; one item per server.
\item $k$ copies of each of the remaining $(n-k)$ items are stored in $k$ servers; one copy of each item per server.
\end{enumerate}
\begin{brtheorem}[{\bf \cite{PaStWe}}]
$N(m+1, k, m) = m+k$.
\label{thm2}
\end{brtheorem}
\par
In this case, for optimal CBC, items are stored in the following way:
\begin{enumerate}[topsep=0pt, label=(\roman*)]
\setlength{\itemsep}{0pt}
\setlength{\parskip}{0pt}
\setlength{\parsep}{0pt}
\setlength{\topsep}{0pt}
\setlength{\partopsep}{0pt}
\setlength{\topskip}{0pt}
\setlength{\headsep}{0pt}
\setlength{\topmargin}{0pt}
\setlength{\itemindent}{7pt}
\setlength{\listparindent}{7pt}
\item Any $m$ of the $m+1$ items are stored in $m$ servers; one item per server.
\item $k$ copies of the remaining item are stored in any $k$ of the $m$ servers.
\end{enumerate}
 Note that this construction is given in \cite{BrKiMeSc} and differs from the construction given in \cite{PaStWe}.
\begin{brtheorem}[{\bf \cite{PaStWe}}]
If $n \geq (k-1) \binom {m}{k-1}$, then $N(n, k, m) = kn - (k-1) \binom{m}{k-1}$.
\label{thm3}
\end{brtheorem}
In this case, optimal CBC is obtained in the following way:
\begin{enumerate}[topsep=0pt, label=(\roman*)]
\setlength{\itemsep}{0pt}
\setlength{\parskip}{0pt}
\setlength{\parsep}{0pt}
\setlength{\topsep}{0pt}
\setlength{\partopsep}{0pt}
\setlength{\topskip}{0pt}
\setlength{\headsep}{0pt}
\setlength{\topmargin}{0pt}
\setlength{\itemindent}{7pt}
\setlength{\listparindent}{7pt}
\item $(k-1) \binom{m}{k-1}$ of the $n$ items are grouped into $\binom{m}{k-1}$ groups of $k-1$ items. These $\binom{m}{k-1}$ groups are stored in $\binom{m}{k-1}$ combinations $k-1$ servers; one group per combination. $k-1$ copies of each of the $k-1$ items of a group are stored in $k-1$ servers of the corresponding combination; one copy of each item per server.
\item For the remaining $n - (k-1) \binom{m}{k-1}$ items, $k$ copies of each are stored in any $k$ of the $m$ servers.
\end{enumerate}
\hspace{1pt}
Note that in \cite{PaStWe}, the authors described the constructions in the setting of ``{\em dual set system}'', which we will describe in the next section.
In \cite{BrKiMeSc} and \cite{BrKiMeSc2}, Brualdi et al. obtained the following optimality result for $n = m+2$ using {\em cocircuit representation} of {\em transversal matroids}.\footnote{In a very recent work Bujt\'{a}s and Tuza studied this case in the framework of hypergraphs. See \cite{BuTu1} for more details.}
\begin{brtheorem}[{\bf\cite{BrKiMeSc}}]
Let $k$ and $m$ be integers with $2 \leq k \leq m$. Then\\
$N(m+2, k, m) = \left \{
\begin{array}{l l}
m+k-2+\lceil 2 \sqrt{k+1} \rceil & \quad \text{if} \quad m+1-k \geq \lceil \sqrt{k+1} \rceil,\\
2m-2 +\lceil 1+\frac{k+1}{m+1-k}\rceil & \quad \text{if} \quad m+1-k < \lceil \sqrt{k+1} \rceil.\\
\end{array} \right.$
\label{thm4}
\end{brtheorem}
\hspace{2pt}
\par
In this article, we obtain following new results for optimal CBCs.
\begin{enumerate}[topsep=1pt]
\setlength{\itemsep}{1pt}
\setlength{\parskip}{0pt}
\setlength{\parsep}{0pt}
\setlength{\topsep}{0pt}
\setlength{\partopsep}{0pt}
\setlength{\topskip}{0pt}
\setlength{\headsep}{0pt}
\setlength{\topmargin}{0pt}

\item By extending a technique of \cite{PaStWe}, we obtain a lower bound on $N(n, k, m)$ for values of $n$ in the range $1\leq n \leq (k-1)\binom{m}{k-1}$.
\item We give explicit construction of optimal CBCs for values of $n$ in the range $\binom{m}{k-2} \leq n \leq (k-1) \binom{m}{k-1}$.
\item Using {\em binary constant weight codes}, we give explicit construction for the range $\binom{m}{k-2}-(m-k+1)A(m, 4, k-3) \leq n \leq \binom{m}{k-2}$ for $k \geq 5$, where $A(m, 4, k-3)$ is the maximum number of codewords of a binary constant weight code of length $m$, weight $k-3$ and Hamming distance $4$. This construction yields optimal CBCs for approximately half of the values of $n$ in this range. For the other half, the construction yields almost optimal CBCs; for these CBCs value of $N$ differs by one from the corresponding value of $N$ given by the lower bound that we have obtained. 
\end{enumerate}
\par
Constructions of (2) and (3), which produce optimal CBCs, show that the lower bound of (1) is best possible for the corresponding ranges and also settle the problem of finding $N(n, k, m)$ for these ranges - partial solution to a problem left open in \cite{PaStWe}.\par
A $c$-uniform $(n, cn, k, m)$-CBC is a CBC where each item is stored in exactly $c$ servers.  In \cite{PaStWe}, the authors gave non-constructive proof of existence of $c$-uniform $(n, cn, k, m)$-CBCs, for which $n$ is $\Omega(m^{\frac{ck}{k-1}-1})$. Using binary constant weight codes, we provide explicit construction  of $c$-uniform $(n, cn, k, m)$-CBCs for $1 \leq \lfloor \frac{k}{2} \rfloor \leq c < k-1$. For sufficiently large $m, c,$ and $k$ such that $c \sim k$ and $c^c \sim m$, our explicit construction is for a value (in asymptotic sense) of $n$ that compares well with the bound $\Omega(m^{\frac{ck}{k-1}-1})$.
\par 
\section{Setting and Preliminaries}
Let $\mathcal{C}$ be an $(n, N, k, m)$-CBC, where $x_1, x_2, \ldots, x_n$ are $n$ items and $s_1, s_2, \ldots, s_m$ are $m$ servers. We represent $\mathcal{C}$ by a {\em set system} $(\mathcal{S, X})$, where $\mathcal{S} = \{s_1, s_2, \ldots, s_m\}$ is the set of $m$ servers and $\mathcal{X} =(X_1, X_2, \ldots, X_n)$ is a collection of $n$ subsets of $\mathcal{S}$, each subset representing one of the $n$ items. If an item $x_j$ is stored in servers $s_{i_1}, s_{i_2}, \ldots, s_{i_l}$, then we represent $x_j$ by a subset $X_j$, where $X_j = \{s_{i_1}, s_{i_2}, \ldots, s_{i_l}\}$, $1 \leq j \leq n$, $\{i_1, i_2, \ldots, i_l\}\subseteq[m]$. For example, let $\mathcal{S} = \{ s_1, s_2, s_3\}$ be the set of servers. Let server $s_1$ contain items $x_1, x_2$, and $x_3$, server $s_2$ contain items $x_1$ and $x_2$, and server $s_3$ contain item  $x_2$. Then the collection $\mathcal{X}$ is $(\{s_1, s_2\}, \{s_1, s_2, s_3\}, \{s_1\})$, where set $\{s_1, s_2\}$ represents item $x_1$ (since it is stored in servers $s_1$ and $s_2$), set $\{s_1, s_2, s_3\}$ represents item $x_2$,  and set $\{s_1\}$ represents item $x_3$. In \cite{PaStWe}, this setting was referred to as {\em dual set system}.\par
Since $\mathcal{C}$ is an $(n, N, k, m)$-CBC, hence total number of items stored in $m$ servers is $N$. So, counting in terms of number of servers that store a particular item, we have in the setting of set system $(\mathcal{S}, \mathcal{X})$ that $\sum_{X \in \mathcal{X}}\lvert X \rvert = N$.\par
Now,  it may be observed that a set of $k$ items $x_{i_1}, x_{i_2}, \ldots, x_{i_k}$ can be retrieved by reading at most one item per server iff there are $k$ distinct servers $s_{j_1}, s_{j_2}, \ldots, s_{j_k} \in \mathcal{S}$ such that $s_{j_r} \in X_{i_r}$ for $1 \leq r \leq k$, which is same as saying that the collection $(X_{i_1}, X_{i_2}, \ldots, X_{i_k})$ has a {\em system of distinct representatives} (SDR). In the previous example, $s_2, s_3, s_1$ is an SDR for the collection $(\{s_1, s_2\}, \{s_1, s_2, s_3\}, \{s_1\})$; items $x_1, x_2, x_3$ are retrieved by reading $x_1$ from server $s_2$, $x_2$ from server $s_3$, and $x_3$ from server $s_1$. Since $\mathcal{C}$ satisfies the condition that any $k$ data items can be retrieved by reading at most one item per server, so $(\mathcal{S}, \mathcal{X})$ satisfies the condition that for any $\{i_1, i_2, \ldots, i_k\} \subseteq [n]$, the collection of sets $(X_{i_1}, X_{i_2}, \ldots, X_{i_k})$ of $\mathcal{X}$\footnote{In this article, by `set of a collection' or `subset of a collection' we will mean set or subset contained in a collection as a member.} has an SDR. {\em Hall's Theorem} provides necessary and sufficient conditions for existence of an SDR for a collection of sets. Below we state the theorem.\par
\begin{theorem}[Hall's Theorem({\bf \cite{Bol}})]
A set system $\mathcal{F} = \{A_1, A_2, \ldots, A_m\}$ has a set of distinct representatives iff\par
\hspace{100pt}$\lvert\bigcup \limits_{i\in S}A_i\rvert \geq \lvert S \rvert$\\
for all $S \subseteq \{1, 2, \ldots, m\}$.
\end{theorem}
The necessary and sufficient condition in Hall's Theorem is known as {\em Hall's condition}. So, for $\mathcal{C}$ to satisfy the condition that any $k$ data items can be retrieved by reading at most one item per server, it is necessary and sufficient that $(\mathcal{S}, \mathcal{X})$ satisfies following restricted form of Hall's condition.
\begin{nhtheorem}
HC1[$k$]: Given any $r$ sets $X_{i_1}, X_{i_2}, \ldots, X_{i_r}$ of $\mathcal{X}$, we have that $\lvert\bigcup \limits_{1\leq j \leq r}X_{i_j}\rvert \geq r$ for all $r$, $1 \leq r \leq k$.
\end{nhtheorem}
In other words, union of any $r$ sets of $\mathcal{X}$ contains at least $r$ elements for $1 \leq r \leq k$. Equivalently, the above condition can also be stated in the following way.
\begin{nhtheorem}
HC2[$k$]: Any $r$ element subset of $\mathcal{S}$ contains at most $r$ sets of $\mathcal{X}$ for all $r$, $0 \leq r \leq k-1$.
\end{nhtheorem}
It is sometimes convenient to use the latter of the above two forms of restricted Hall's condition. In this article we will use both forms. Next, we introduce following notations to conveniently express sub-conditions of HC1[$k$] and HC2[$k$].
\begin{enumerate}[topsep=0pt, label=(\roman*)]
\setlength{\itemsep}{0pt}
\setlength{\parskip}{0pt}
\setlength{\parsep}{0pt}
\setlength{\topsep}{0pt}
\setlength{\partopsep}{0pt}
\setlength{\topskip}{0pt}
\setlength{\headsep}{0pt}
\setlength{\topmargin}{0pt}
\item For set system $(\mathcal{S}, \mathcal{X})$ and a subcollection $\mathcal{Y} \subseteq \mathcal{X}$, we denote by HC1($\mathcal{Y}$) the condition that $\lvert\bigcup\limits_{X \in \mathcal{Y}} X \rvert \geq \lvert \mathcal{Y} \rvert$. So, for given $k$, $(\mathcal{S}, \mathcal{X})$ satisfies HC1[$k$] iff for all $\mathcal{Y} \subseteq \mathcal{X}$ with $1 \leq \lvert \mathcal{Y} \rvert \leq k$, HC1($\mathcal{Y}$) is satisfied.
\item For set system $(\mathcal{S}, \mathcal{X})$ and a subset $X \subseteq \mathcal{S}$ with $\lvert X \rvert = i$, we denote by HC2($X$) the condition that $X$ contains at most $i$ sets of $\mathcal{X}$. So, for given $k$, $(\mathcal{S}, \mathcal{X})$ satisfies HC2[$k$] iff for all $X \subseteq \mathcal{S}$ with $0 \leq \lvert X\rvert \leq k-1$, HC2($X$) is satisfied.
\end{enumerate}
\par
Also, to simplify the presentation, we identify a CBC with the corresponding set system. For example, we say `$(\mathcal{S}, \mathcal{X})$ is an $(n, N, k, m)$-CBC' rather than `$(\mathcal{S}, \mathcal{X})$ corresponds to (or represents) an $(n, N, k, m)$-CBC'. \par
At this point, it is interesting to note that for a set system $(\mathcal{S}, \mathcal{X})$  the sub-conditions HC2($X$), $X \subseteq \mathcal{S}$, are independent, in the sense that for a given subset $X$ of $\mathcal{S}$, satisfaction of every other HC2($Y$), $Y \subseteq \mathcal{S}$, $Y \neq X$, does not necessarily imply satisfaction of HC2($X$). For example, let $\mathcal{S} = \{s_1, s_2, s_3\}$ and let $\mathcal{X} = (\{s_1, s_2\}, \{s_1, s_2\}, \{s_1, s_2\})$. Then it can be verified that only HC2($\{s_1, s_2\}$) is violated.\par
Now, for $(n, N, k, m)$-CBC $(\mathcal{S}, \mathcal{X})$ and for each $i$, $1 \leq i \leq k-1$, we use the method of {\em double-counting} to `combine' $\binom {m}{i}$ sub-conditions HC2($X$), $X \subseteq \mathcal{S}, \lvert X \rvert = i$, into an inequality in the following way: we form an $\binom{m}{i} \times n$ matrix $M^i$, whose rows are labeled by all the $i$-subsets (i.e., $i$ element subsets, for convenience we will use this short form) of $\mathcal{S}$, and columns are labeled by the $n$ sets of $\mathcal{X}$. $(r, s)$-th entry of $M^i$ is $1$ if the $s$-th set of $\mathcal{X}$ is contained in the $r$-th $i$-subset of $\mathcal{S}$, otherwise it is $0$. Let $A_j$ denote the number of $j$-sets of $\mathcal{X}$, $1 \leq j \leq k-1$. Next, we count the number of $1$s in $M^i$ in two ways. Since HC2($X$) is satisfied for all $X \subseteq \mathcal{S}, \lvert X \rvert = i$, so each row has at most $i$ $1$s; hence, counting row-wise, there are at most $i\binom{m}{i}$ $1$s in $M^i$. A column labeled by $X' \in \mathcal{X}$ with $1 \leq \lvert X' \rvert = j \leq i$, has exactly $\binom{m-j}{i-j}$ $1$s, and has none if $j > i$. Hence, counting column-wise, there are exactly $\sum_{j=1}^{i}\binom{m-j}{i-j}A_j$ $1$s in $M^i$. Comparing these two numbers, we get the following inequality.\par
\begin{equation}
\sum_{j=1}^{i}\binom{m-j}{i-j}A_j \leq i\binom{m}{i}, \hspace{1 in} 1 \leq i \leq k-1.
\label{systineq1}
\end{equation}
So, considering every $i$ in the range $1 \leq i \leq k-1$, we get $k-1$ inequalities, each of which is satisfied by the $(n, N, k, m)$-CBC $(\mathcal{S}, \mathcal{X})$. Here it is important to note that $k-1$ inequalities, unlike the conditions HC2($X$), $X \subseteq \mathcal{S}$ (from which they are derived), are not mutually independent. In fact, we show in the next lemma that if $(i+1)$-th  inequality is satisfied, then $i$-th inequality is also satisfied. 
\begin{lemma}
For $m \geq 3$ and $1 \leq i \leq m-2$ if $\sum_{j=1}^{i+1}\binom{m-j}{i-j+1}A_j \leq (i+1) \binom{m}{i+1}$, then $\sum_{j=1}^{i}\binom{m-j}{i-j}A_j \leq i\binom{m}{i}$.
\end{lemma}
\pf We prove the contrapositive, i.e., we show that if $\sum_{j=1}^{i}\binom{m-j}{i-j}A_j > i \binom{m}{i}$, then $\sum_{j=1}^{i+1}\binom{m-j}{i-j+1}A_j > (i+1)\binom{m}{i+1}$.\\
Now, $ \sum_{j=1}^{i+1}\binom{m-j}{i-j+1}A_j = A_{i+1}+(m-i)\sum_{j=1}^{i}\dfrac{\binom{m-j}{i-j}A_j}{i-j+1} \geq A_{i+1}+\dfrac{m-i}{i} \sum_{j=1}^{i}\binom{m-j}{i-j}A_j.$\\
Since we assumed $\sum_{j=1}^{i}\binom{m-j}{i-j} A_j > i \binom{m}{i}$, we have from the above that\\
$\sum_{j=1}^{i+1}\binom{m-j}{i-j+1}A_j  > A_{i+1}+(m-i)\binom{m}{i} \geq (i+1)\binom{m}{i+1}.$\\
Last step follows from the fact that $A_{i+1} \geq 0 $ and $(m-i)\binom{m}{i} = (i+1)\binom{m}{i+1}$.
\qed\\
\\
Hence, it follows that if the $(k-1)$-th inequality, i.e.,  
\begin{equation}
\sum_{j=1}^{k-1}\binom{m-j}{k-j-1}A_j \leq (k-1)\binom{m}{k-1}, \hspace{1 in}
\label{mainineq1}
\end{equation}
is satisfied, then the remaining $k-2$ inequalities are also satisfied. That is, as necessary conditions, the other $k-2$ inequalities are redundant with respect to \ref{mainineq1}, and excluding these $k-2$ inequalities from further considerations, we will only use the fact that an $(n, N, k, m)$-CBC $(\mathcal{S}, \mathcal{X})$ satisfies \ref{mainineq1} as necessary condition. Note that in \cite{PaStWe}, the authors obtained inequality \ref{mainineq1} in the proof of Theorem \ref{thm3} by considering the case of $i = k-1$ only. What we have also shown here is that the other $k-2$ inequalities, obtained by considering HC2($i$), $1 \leq i \leq k-2$, in a similar way, are redundant; a fact which was not observed in \cite{PaStWe} and was not very evident in the first place.\par
Inequality \ref{systineq1} is a `tight' necessary condition for CBCs for wide ranges of values of $n$, in the sense that for these ranges of values of $n$ there are CBCs that just satisfy this inequality (i.e., satisfy with equality), and these are the optimal CBCs for the corresponding ranges (e.g., optimal CBCs constructed in \cite{PaStWe} for $n \geq (k-1) \binom{m}{k-1}$). In the next section, we implicitly demonstrate this `tightness' by using \ref{mainineq1} to obtain a lower bound on $N(n, k, m)$ for $1 \leq n \leq (k-1) \binom{m}{k-1}$, and then constructing optimal CBCs (i.e., CBCs that meet this lower bound) for a sub-range of $1 \leq n \leq (k-1) \binom{m}{k-1}$.\par 
As an immediate corollary of this inequality, we state the following theorem of \cite{PaStWe}.\par 
\begin{theorem} [{\bf \cite{PaStWe}}]
$n(m, c, k) \leq \dfrac{(k-1)\binom{m}{c}}{\binom{k-1}{c}}$, where $n(m, c, k)$  is the maximum value of $n$ such that there exists a $c$-uniform $(n, cn, k, m)$-CBC.
\label{unibound}
\end{theorem}
$c = 1, k-2,$ and $k-1$ are the only values, known so far, for which there exists $c$-uniform $(n, cn, k, m)$-CBCs with $n$ given by the expression of Theorem \ref{unibound}.\par
\section{Our Results}
\subsection{Lower bound on $N(n,k,m)$ for $1 \leq n \leq (k-1)\binom{m}{k-1}$}
Here we obtain a lower bound on $N(n, k, m)$ for $1 \leq n \leq (k-1)\binom{m}{k-1}$. The proof is divided into three steps. First, we prove a useful inequality in Lemma \ref{ineq1}. In Lemma \ref{lem1}, we use inequality \ref{systineq1} to get a relationship between $N$ and $\frac{(k-1)\binom{m}{c}}{\binom{k-1}{c}}$ for any $c$ such that $1 \leq c \leq  k-1$. From this relationship we get different estimates of lower bound on $N(n, k, m)$ for different values of $c$, $1 \leq c \leq k-1$. Our use of \ref{systineq1} is same as in the proof of Theorem \ref{thm3} of \cite{PaStWe}. There, the authors obtained lower bound on $N(n, k, m)$ for $n \geq (k-1) \binom{m}{k-1}$; they set-up relationship between $N$ and $(k-1)\binom{m}{k-1}$ (value of $\frac{(k-1)\binom{m}{c}}{\binom{k-1}{c}}$ at $c = k-1$). Here we generalize (using Lemma \ref{ineq1}) their approach for any $c$ in the stated range. Finally, in Theorem \ref{th2}, we find $c$ that gives best estimate for lower bound on $N(n, k, m)$. \par

\begin{lemma}
Let $1 \leq c < k \leq m$ and $0 \leq i \leq k-1$. Then $\dfrac{\binom{m-i}{k-1-i}}{\binom{m-c}{k-1-c}} -1 \geq \dfrac{(m-k+1)(c-i)}{k-c}$.
\label{ineq1}
\end{lemma}
\pf Notice that both sides are equal for $i =c$ and $i = c-1$, and both sides decrease as $i$ goes from $0$ to $k-1$. Hence, it is sufficient to show that difference between l.h.s. values for $i-1$ and $i$ is greater than or equal to $\dfrac{m-k+1}{k-c}$ for $2 \leq i \leq c-1$, and is less than or equal to $\dfrac{m-k+1}{k-c}$ for $c+1 \leq i \leq k-1$.\\
\vspace{5pt}
Now,  $\dfrac{\binom{m-i+1}{k-i}}{\binom{m-c}{k-c -1}} - \dfrac{\binom{m-i}{k-i-1}}{\binom{m-c}{k-c-1}} = \dfrac{\frac{m-k+1}{k-i} \binom{m-i}{k-i-1}}{\binom{m-c}{k-c-1}} = \dfrac{\frac{m-k+1}{k-c} \binom{m-i}{k-i}}{\binom{m-c}{k-c}}$. \\
Here we note that when $c > i$, we have $\binom{m-i}{k-i}\binom{k-i}{c-i}= \binom{m-i}{c-i}\binom{m-c}{k-c}$.\\
And when $i>c$, we have $\binom{m-c}{k-c}\binom{k-c}{i-c}= \binom{m-c}{i-c}\binom{m-i}{k-i}$.\newline
\vspace{5pt}
\Big[ In the above two cases we have used the identity $\binom{x}{y}\binom{y}{z} = \binom {x}{z} \binom{x-z}{y-z}$, for $x \geq y \geq z \geq 0$. \Big]\\
\vspace{5pt}
$\text{Hence,} \; \dfrac{\frac{m-k+1}{k-c} \binom{m-i}{k-i}}{\binom{m-c}{k-c}}= \left\{
  \begin{array}{l l}
    \dfrac{\frac{m-k+1}{k-c} \binom{m-i}{c-i}}{\binom{k-i}{c-i}} \geq \dfrac{m-k+1}{k-c} & \quad \text{when $c > i$},\\
     \dfrac{\frac{m-k+1}{k-c} \binom{k-c}{i-c}}{\binom{m-c}{i-c}} \leq \dfrac{m-k+1}{k-c}& \quad \text{when $ i > c$
.}\\
  \end{array} \right.$
\qed
\\

Let us use the notation $U_{m, k, c}$ for the expression $\frac{(k-1)\binom{m}{c}}{\binom{k-1}{c}}$. Note that $U_{m, k, c}$ may not be an integer for given values of $m, c,$ and $k$ and that $U_{m, k, c} < U_{m, k, c'}$ for $c < c'$.\par
\begin{lemma}
Let $(\mathcal{S}, \mathcal{X})$ be an $(n, N, k, m)$-CBC and $1 \leq c \leq k-1$. Then $N \geq nc -  \frac{(k-c)(U_{m,k,c}-n)}{m-k+1}+\frac{(k-c)(m-k)}{m-k+1}A_k$, where $A_k$ is the number of $k$-sets of $\mathcal{X}$.
\label{lem1}
\end{lemma}
\pf Since $(\mathcal{S}, \mathcal{X})$ satisfies HC1[$k$], hence it is sufficient for a set of $\mathcal{X}$ to be of size $k$, in the sense that if a subcollection $\mathcal{Y} \subseteq \mathcal{X}$ with $1 \leq \lvert \mathcal{Y} \rvert \leq k$ contains a $k$-set, then it always satisfies HC1($\mathcal{Y}$). Hence, without loss of generality, we assume that each set of $\mathcal{X}$ is of cardinality at most $k$. Let $A_i$ be the number of $i$-sets of $\mathcal{X}$. Then we have the following equation.
\begin{equation}
\sum_{i=1}^{k}A_i = n.
\label{eq1}
\end{equation}
As a necessary condition $(\mathcal{S}, \mathcal{X})$ also satisfies the following inequality.
\begin{equation}
\sum_{i=1}^{k-1}\binom{m-i}{k-i-1}A_i \leq (k-1)\binom{m}{k-1}.
\label{eq2}
\end{equation}
Dividing both sides of \ref{eq2} by $\binom{m-c}{k-c-1}$ and then subtracting \ref{eq1}, we get \\
\begin{equation}
\sum_{i = 1}^{k-1} (\dfrac{\binom{m-i}{k-i-1}}{\binom{m-c}{k-c-1}}-1) A_i - A_k \leq U_{m, k, c}-n.
\label{eq3}
\end{equation}
Employing Lemma \ref{ineq1} to \ref{eq3}, we get
\begin{equation}
\sum_{i = 1}^{k-1} (c-i)A_i \leq \frac{(k-c)(U_{m, k, c}+A_k-n)}{m-k+1}.
\label{eq4}
\end{equation}
Now,\\
\begin{equation}
N = \sum_{i=1}^{k} i A_i\\
= nc-\sum_{i=1}^{k}(c-i)A_i.
\end{equation}
Using \ref{eq4}, we get
\begin{equation}
N \geq nc -  \frac{(k-c)(U_{m,k,c}-n)}{m-k+1}+\frac{(k-c)(m-k)}{m-k+1}A_k.
\label{lemineq}
\end{equation}
\qed
\\
\\
\begin{theorem}
Let $1\leq n \leq (k-1) \binom{m}{k-1}$, $1 \leq c \leq  k-1$, and $c$ be the least integer such that $n \leq U_{m, k, c}$. Then $ nc - \left\lfloor \frac{(k-c)(U_{m,k,c} -n)}{m-k+1} \right\rfloor$ is a lower bound for  $N(n, k, m)$.
\label{th2}
\end{theorem}
\pf Let $(\mathcal{S}, \mathcal{X})$ be an optimal $(n, N, k, m)$-CBC (i.e., for it  $N = N(n, k, m)$) and $A_k$ be the number of $k$-sets of $\mathcal{X}$. Then for any $i$, $1 \leq i \leq k-1$, we have from relation \ref{lemineq} of Lemma \ref{lem1} that $N(n, k, m) \geq ni -  \frac{(k-i)(U_{m, k, i} - n)}{m-k+1} + \frac{(k-i)(m-k)}{m-k+1}A_k$. Since $A_k \geq 0$, we have $N(n, k, m) \geq ni -  \frac{(k-i)(U_{m, k, i} - n)}{m-k+1}$.  Let us denote $ni - \frac{(k-i)(U_{m, k, i}-n)}{m-k+1}$ by $b(n, k, m, i)$. Next, we find $i$ which maximizes $b(n, k, m, i)$. For $1\leq i \leq k-2$, we have\\
\begin{align*}
b(n, k, m, i) - b(n, k, m, i+1) &= ni - \frac{(k-i)(U_{m, k, i}-n)}{m-k+1} -n(i+1) + \frac{(k-i-1)(U_{m, k, i+1}-n)}{m-k+1}\\
&= \frac{(k-i-1)(U_{m, k, i+1}- U_{m, k, i})}{m-k+1} - \frac{U_{m, k, i}-n}{m-k+1} - n\\
&= U_{m, k, i} -n - \frac{U_{m, k, i}-n}{m-k+1} \hspace{5pt} \left[\text{using} \quad U_{m, k, i+1} - U_{m, k, i} = \frac{(m-k+1)U_{m, k, i}}{k-i-1}\right]\\
&= \frac{(m-k)(U_{m, k, i}-n)}{m-k+1}.
\end{align*}
Let $c$ be such that $U_{m,k,1}< U_{m, k, 2} < \ldots <n \leq U_{m, k, c} < U_{m, k, c+1} \ldots <U_{m, k, k-1}, 1 \leq c \leq k-1$, then using above relation we have $b(n, k, m, 1) < b(n, k, m, 2) \ldots < b(n, k, m, c) \geq b(n, k, m, c+1) \ldots > b(n,k, m, k-1)$.\\
Since $N(n, k, m)$ is an integer, we have $N(n, k, m) \geq \lceil b(n, k, m, c)\rceil \geq  nc - \left\lfloor \frac{(k-c)(U_{m, k, c}-n)}{m-k+1} \right\rfloor$ and $nc - \left\lfloor \frac{(k-c)(U_{m, k, c}-n)}{m-k+1} \right\rfloor$ is a lower bound on $N(n, k, m)$.
\qed
\subsection{Construction of optimal CBCs for the range $\binom{m}{k-2} \leq n \leq (k-1)\binom{m}{k-1}$}
Let $\mathcal{S}$ be the set of servers, where $\lvert \mathcal{S} \rvert = m$. Avoiding the trivial case of $k = 2$, for which the range becomes $1 \leq n \leq m$, we consider cases where $m\geq k \geq 3$. Roughly, the construction is as follows. We start with a CBC $(\mathcal{S}, \mathcal{X}_i)$, in which $\mathcal{X}_i$ is a collection of $(k-1)$-subsets of $\mathcal{S}$. We also take an auxiliary collection $\mathcal{X}_a$ of distinct $(k-2)$-subsets\footnote{Here `distinct' means that the subsets contain different elements, i.e., they are distinct as subsets of $\mathcal{S}$. For the rest of this article, this interpretation will be assumed.} of $\mathcal{S}$. From $\mathcal{X}_i$ we systematically delete $(k-1)$-sets and add to it $(k-2)$-sets from $\mathcal{X}_a$ to get the final collection $\mathcal{X}$. Below we describe the construction in more detail.\par
\cn As initial collection, we take the collection $\mathcal{X}_i$ of sets of an optimal $(n, N, k, m)$-CBC $(\mathcal{S}, \mathcal{X}_i)$, where $n = U_{m, k, k-1}=(k-1)\binom{m}{k-1}$\footnote{So far, there is only one optimal collection known for $n=U_{m, k, k-1}$; hence, we write `the collection' here. We do the same in the next construction for $n = U_{m, k, k-2}$.}. In this collection, there are $k-1$ copies of each of the $(k-1)$-subsets of $\mathcal{S}$. \par
For the CBC to be constructed, we have $U_{m, k, k-2} \leq n \leq U_{m, k, k-1}$. Hence, $0 \leq U_{m, k, k-1}-n \leq (m-k+1)\binom{m}{k-2}$. The auxiliary collection $\mathcal{X}_a$ contains any $\left\lceil \frac{U_{m, k, k-1}-n}{m-k+1}\right\rceil$ distinct $(k-2)$-subsets of $\mathcal{S}$. This is possible given the range of values of $U_{m, k, k-1} - n$. Next, we do the following $\left\lfloor \frac{U_{m, k, k-1} - n}{m-k+1} \right\rfloor$ times.
\begin{enumerate}
\item Select a $(k-2)$-set from $\mathcal{X}_a$ and delete one copy of each of its $m-k+2$ supersets from $\mathcal{X}_i$. For each selected $(k-2)$-set of $\mathcal{X}_a$, we can always delete one copy of each of its $m-k+2$ supersets from $\mathcal{X}_i$ irrespective of previous deletions. This is because there are $k-1$ copies of each of the $(k-1)$-subsets of $\mathcal{S}$ in the initial collection $\mathcal{X}_i$. So, for a $(k-1)$-subset of $\mathcal{X}_i$, its $k-1$ copies may be assumed to be assigned to its $k-1$ distinct $(k-2)$-subsets; one copy per subset. Therefore, for a $(k-2)$-set of $\mathcal{X}_a$ there corresponds a copy of each of its $m-k+2$ supersets in $\mathcal{X}_i$.
\item Add the $(k-2)$-set to the collection $\mathcal{X}_i$ and delete it from the auxiliary collection $\mathcal{X}_a$.
\end{enumerate}
Finally, if $(m-k+1) \nmid (U_{m, k, k-1} - n)$, then for the remaining $(k-2)$-set of $\mathcal{X}_a$, delete one copy of each of its $(U_{m, k, k-1}-n) - \left\lfloor \frac{U_{m, k, k-1}-n}{m-k+1}\right\rfloor(m-k+1)$ supersets from $\mathcal{X}_i$.\par 
In the end, we get the final collection $\mathcal{X}$ of $n$ subsets of $\mathcal{S}$. Before proving its correctness, we give an example to illustrate the construction.
\begin{example}
Let us take $m = 6, k = 4, n = 43$ and $\mathcal{S}=\{s_1, s_2, s_3, s_4, s_5, s_6\}$. Hence, $U_{m,k,k-1} = 60$, and the initial collection $\mathcal{X}_i$ contains $k-1 = 3$ copies of each of the $20$ $3$-subsets of $\mathcal{S}$.  $\left\lceil \frac{U_{m, k, k-1} - n}{m-k+1} \right\rceil = 6$, and the auxiliary collection $\mathcal{X}_a$ contains $6$ $2$-subsets of $\mathcal{S}$; let it be $(\{s_1, s_2\}, \{s_2, s_3\}, \{s_3, s_4\}, \{s_4, s_5\}, \{s_5, s_6\}, \{s_1, s_6\})$. \par
For step (i), we select subset $\{s_1, s_2\}$ of $\mathcal{X}_a$, delete one copy of each of its $m-k+2=4$ supersets (i.e., $\{s_1, s_2, s_3\}, \{s_1, s_2, s_4\}, \{s_1, s_2, s_5\}, \{s_1, s_2, s_6\}$) from $\mathcal{X}_i$; add the subset $\{s_1, s_2\}$ to $\mathcal{X}_i$ and delete it from $\mathcal{X}_a$. We repeat these steps for $4$ other subsets (let us take $\{s_2, s_3\}, \{s_3, s_4\}, \{s_4, s_5\}, \{s_5, s_6\}$) of $\mathcal{X}_a$. \par
Finally, for the remaining subset $\{s_1, s_6\}$, we delete two of its supersets $\{s_1, s_2, s_6\}$ and $\{s_1, s_3, s_6\}$ from collection $\mathcal{X}_i$. Table \ref{tab1} shows the final collection $\mathcal{X}$.\par

\begin{table}[h]
\caption{Final collection $\mathcal{X}$ of Example \ref{ex1}}
\begin{center}
\begin{tabular}{|l|c|l|c|}

\hline
Subset & Number of copies & Subset & Number of copies\\
\hline
$\{s_1, s_2, s_3\}$ & $1$ & $\{s_2, s_3, s_6\}$ & $2$\\
$\{s_1, s_2, s_4\}$ & $2$ & $\{s_2, s_4, s_6\}$ & $3$\\
$\{s_1, s_2, s_5\}$ & $2$ & $\{s_2, s_5, s_6\}$ & $3$\\
$\{s_1, s_2, s_6\}$ & $2$ & $\{s_3, s_4, s_5\}$ & $1$\\
$\{s_1, s_3, s_4\}$ & $2$ & $\{s_3, s_4, s_6\}$ & $2$\\
$\{s_1, s_3, s_5\}$ & $3$ & $\{s_3, s_5, s_6\}$ & $2$\\
$\{s_1, s_3, s_6\}$ & $2$ & $\{s_4, s_5, s_6\}$ & $1$\\
$\{s_1, s_4, s_5\}$ & $2$ & $\{s_1, s_2\}$ & $1$\\
$\{s_1, s_4, s_6\}$ & $3$ & $\{s_2, s_3\}$ & $1$\\
$\{s_1, s_5, s_6\}$ & $2$ & $\{s_3, s_4\}$ & $1$\\
$\{s_2, s_3, s_4\}$ & $1$ & $\{s_4, s_5\}$ & $1$\\
$\{s_2, s_3, s_5\}$ & $2$ & $\{s_5, s_6\}$ & $1$\\
\hline
\end{tabular}
\end{center}
\label{tab1}
\end{table}
\label{ex1}
\end{example}
\pocr
 Note that sets of $\mathcal{X}$ are of cardinality $k-1$ and $k-2$, and the $(k-2)$-sets are distinct. In order to prove correctness of the construction, we show that $(\mathcal{S},\mathcal{X})$ satisfies HC1[$k$], i.e., for all subcollections $\mathcal{Y} \subseteq \mathcal{X}$ with $1 \leq \lvert \mathcal{Y}\rvert \leq k$, HC1($\mathcal{Y}$) is satisfied.
\begin{enumerate}
\item {\bf HC1($\mathcal{Y}$) for $\mathcal{Y} \subseteq \mathcal{X}$ with $1 \leq \lvert \mathcal{Y} \rvert \leq k-1$:} For any subcollection $\mathcal{Y} \subseteq \mathcal{X}$ with $1 \leq \lvert \mathcal{Y} \rvert \leq k-2$, HC1($\mathcal{Y}$) is trivially satisfied. Since union of two distinct $(k-2)$-sets contains at least $k-1$ elements and we are considering cases where $k-1 \geq 2$, it follows that HC1($\mathcal{Y}$) is also satisfied for a subcollection $\mathcal{Y} \subseteq \mathcal{X}$ with $\lvert \mathcal{Y} \rvert = k-1$.
\item {\bf HC1($\mathcal{Y}$) for $\mathcal{Y} \subseteq \mathcal{X}$ with $\lvert \mathcal{Y} \rvert = k$:} Consider any subcollection $\mathcal{Y} \subseteq \mathcal{X}$ such that $\lvert \mathcal{Y} \rvert = k$. One of the following applies for $\mathcal{Y}$. 
\begin{enumerate}
\item All the sets of $\mathcal{Y}$ are $(k-2)$-sets. Since a $(k-1)$-set contains $k-1$ distinct $(k-2)$-subsets, therefore union of $k$ distinct $(k-2)$-sets contains at least $k$ elements. So, HC1($\mathcal{Y}$) is satisfied.
\item $\mathcal{Y}$ has one or more copies of a $(k-1)$-set $X$. In this case, observe that $\mathcal{Y}$ has a set $Y$ such that $Y\subsetneq X$. This is because, in the initial collection $\mathcal{X}_i$ there are $k-1$ copies of $X$ (including itself), and during construction a subset of $X$ is added to $\mathcal{X}_i$ after deleting one copy of  $X$. So, in the final collection $\mathcal{X}$, there are exactly $k-2$ other sets $X'$ such that $X'\subseteq X$.\\
So, we have $\lvert X \cup Y\rvert \geq k$. Hence, HC1($\mathcal{Y}$) is satisfied. \qed
\end{enumerate}
\end{enumerate}
\par
 So, $(\mathcal{S}, \mathcal{X})$ is an $(n, N, k, m)$-CBC, where $N = \sum_{X \in \mathcal{X}} \lvert X \rvert = n(k-1) - \left\lfloor \frac{U_{m,k,k-1} - n}{m-k+1} \right\rfloor$. Hence, following Theorem \ref{th2}, it is an optimal CBC. Therefore, we have proved the following.
\begin{theorem}\footnote{In a very recent and independent work Bujt\'{a}s and Tuza also proved this result. See \cite{BuTu2} for this and more results.}
Let $\binom{m}{k-2} \leq n \leq (k-1)\binom{m}{k-1}$. Then $N(n, k, m) = n(k-1) - \left\lfloor \frac{(k-1)\binom{m}{k-1}-n}{m-k+1} \right\rfloor$. 
\end{theorem}
\subsection{Construction of optimal and almost optimal CBCs for the range $\binom{m}{k-2}-(m-k+1)A(m, 4, k-3) \leq n \leq \binom{m}{k-2}$ for $k \geq 5$: Construction using Binary Constant Weight Codes}
Before describing the construction, we briefly state relevant results of binary constant weight codes. See \cite{HufPle} or any standard text on coding theory for basic details and terminology related to codes.\par
Let $\mathcal{S}$ be an $l$-set $\{s_1, s_2, \ldots, s_l\}$. For $I \subseteq \mathcal{S}$, {\em characteristic vector} of $I$ is the vector $\chi_{I} = (c_1, c_2, \ldots, c_l) \in \mathbb{F}_{2}^{l}$ such that $c_i = 1$ iff $s_i \in I$, $1 \leq i \leq l$, where $\mathbb{F}_2$ is the finite field of order $2$. So, a subset of a set can be naturally identified with its characteristic vector.\par
A binary constant weight code is a nonlinear code over $\mathbb{F}_2$, whose every codeword has same weight. In order to apply the results of binary constant weight codes into the setting of set system, we view codewords as characteristic vectors of subsets. A $w$-subset of an $l$-set is identified with a codeword of length $l$ and weight $w$, where the codeword is the characteristic vector of the subset. Thus, if distance between two codewords is $d$, then symmetric difference between the two corresponding subsets is also $d$; we say that such a pair of subsets is $d$ {\em distance apart}. Here it may be observed that if two sets have same cardinality, then cardinality of their symmetric difference is even; or equivalently, distance between two codewords of same weight is even. \footnote{Due to this, distance between two codewords of a binary constant weight code is commonly given in terms of $2d$ rather than $d$.} \par
In the context of CBCs, we are concerned with construction and bound, especially lower bound, on the size of binary constant weight codes. Construction of binary constant weight codes, although very interesting, will not be discussed. But to give an approximate idea of the range of values $n$ covered by our construction, we state two general lower bounds obtained in \cite{GraSlo}.\par
Let $A(n, 2d, w)$ denote maximum number of codewords of a binary constant weight code of length $n$, weight $w$ and minimum distance $2d$ over field $\mathbb{F}_2$. In ~\cite{GraSlo}, Graham and Sloane gave an elegant construction of binary constant weight codes with minimum distance $4$, which leads to the following lower bound on the maximum number of codewords.
\begin{theorem}[{\bf\cite{GraSlo}}]
$A(n, 4, w) \geq \dfrac{1}{n}\dbinom{n}{w}.$
\label{bound4}
\end{theorem}
They gave another construction for arbitrary $d$, and hence the following lower bound.
\begin{theorem}[{\bf \cite{GraSlo}}]
Let $q$ be a prime power such that $q \geq n$. Then \\
$A(n, 2d, w) \geq \dfrac{1}{q^{d-1}} \dbinom{n}{w}.$
\label{boundd}
\end{theorem}
This, along with Johnson's upper bound on the size of binary constant weight codes, results in the following asymptotic estimate of $A(n, 2d, w)$.
\begin{theorem}[{\bf\cite{GraSlo} }]
$\dfrac{n^{(w-d+1)}}{w!} \lesssim A(n, 2d, w) \lesssim \dfrac{(d-1)! n^{(w-d+1)}}{w!},$ for $w$ fixed as $n\to\infty.$
\label{approx}
\end{theorem}
\begin{remark}
For $d=2$ the above implies $A(n, 4, w) \sim \dfrac{n^{(w-d+1)}}{w!}.$
\end{remark}
There have been various improvements (in specific cases) of the bounds in Theorem \ref{bound4} and Theorem \ref{boundd}. For example, in ~\cite{GraSlo}, the authors themselves gave a construction based on sets with distinct sums, and the lower bound, thus obtained, is better than that of Theorem \ref{boundd} for lower values of $n$. 
For related results and improvements of the bound in Theorem \ref{bound4} see \cite{Klo},\cite{VanEtz}, and a more recent \cite{BroEtz} (and references given there).  For various constructions and bounds of constant weight codes see \cite{BroSheSloSmi}, \cite{AgrVarZeg}, and \cite{SmiHugPer} (and references given there). \par
\cn Our overall construction procedure for this range is similar to our previous construction; although with a different initial collection ($\mathcal{X}_i$) and auxiliary collection ($\mathcal{X}_a$). As in the previous case, we take $\mathcal{S}$ to be the set of servers, $\lvert \mathcal{S} \rvert = m$. We take the initial collection to be the collection $\mathcal{X}_i$ of sets of an optimal $(n, N, k, m)$-CBC $(\mathcal{S}, \mathcal{X}_i)$, where $n = U_{m, k, k-2}=\binom{m}{k-2}$. This collection consists of all the $(k-2)$-subsets of $\mathcal{S}$.  \par 
For the CBC to be constructed, we have $U_{m, k, k-2} - (m-k+1) A(m, 4, k-3) \leq n \leq U_{m, k, k-2}$. Hence, $0 \leq U_{m, k, k-2} - n \leq (m-k+1)A(m, 4, k-3)$. However, unlike in the previous construction, choice of auxiliary collection ($\mathcal{X}_a$) of sets is not arbitrary. In this case, we take $\mathcal{X}_a$ to be a collection of $\left\lceil \frac{U_{m, k, k-2}-n}{m-k+1}\right\rceil$ distinct $(k-3)$-subsets (of $\mathcal{S}$) which are mutually minimum $4$ distance apart. Note that choice of the $(k-3)$-sets of $\mathcal{X}_a$ is guided by codewords of corresponding binary constant weight codes, which is possible for the range of values of $U_{m,k,k-2} - n$. Next, we do the following $\left\lfloor\frac{U_{m, k, k-2} - n}{m-k+1}\right\rfloor$ times.
\begin{enumerate}
\item Select a $(k-3)$-set from $\mathcal{X}_a$ and delete each of its $m-k+3$ supersets from $\mathcal{X}_i$. This can be done for each $(k-3)$-set of $\mathcal{X}_a$ irrespective of all the previous deletions. This is because, $(k-3)$-sets of $\mathcal{X}_a$ are mutually minimum $4$ distance apart. Hence, union of any two subsets of $\mathcal{X}_a$ contains at least $k-1$ elements. So two $(k-3)$-sets of $\mathcal{X}_a$ can not have same $(k-2)$-superset in $\mathcal{X}_i$. 
\item Delete the $(k-3)$-set from $\mathcal{X}_a$ and add two copies of the set to $\mathcal{X}_i$.
\end{enumerate}
Finally, if $(m-k+1) \nmid (U_{m, k, k-2} - n)$, then for the remaining $(k-3)$-set of $\mathcal{X}_a$, delete its $(U_{m, k, k-2} - n) - \left\lfloor \frac{U_{m, k, k-2}- n}{m-k+1}\right\rfloor(m-k+1)$ supersets from $\mathcal{X}_i$.\par
In the end, we get the final collection $\mathcal{X}$ of $n$ subsets of $\mathcal{S}$.\par
\pocr Note that the sets of the final collection $\mathcal{X}$ are of cardinality $k-2$ and $k-3$, and there are exactly two copies of a $(k-3)$-set and the $(k-2)$-sets are all distinct. We prove correctness of the construction by showing that $(\mathcal{S}, \mathcal{X})$ satisfies HC2[$k$], i.e., HC2($X$) is satisfied for all $X \subset \mathcal{S}$ with $0 \leq \lvert X \rvert \leq k-1$ .
\begin{enumerate}
\item {\bf HC2($X$) for $X \subset \mathcal{S}$ with $0 \leq \lvert X \rvert \leq k-2$:} HC2($X$) is trivially satisfied for all $X \subset \mathcal{S}$ with $0 \leq \lvert X \rvert \leq k-4$. Since there are exactly two copies of a $(k-3)$-set in $\mathcal{X}$, HC2($X$) is also satisfied for all $X \subset \mathcal{S}$ with $\lvert X \rvert = k-3$, for $k\geq 5$. Now, two $(k-2)$-sets of $\mathcal{X}_a$ are at least distance $4$ apart, hence their union contains at least $k-1$ elements. Following the construction, union of a $(k-3)$-set and a $(k-2)$-set of $\mathcal{X}$ contains at least $k-1$ elements. Hence, union of more than two sets of $\mathcal{X}$ contains at least $k-1$ elements. Therefore, HC2($X$) is satisfied for all $X \subset \mathcal{S}$ with $\lvert X \rvert = k-2$, for $k \geq 5$.
\item {\bf HC2($X$) for $X \subset \mathcal{S}$ with $\lvert X \rvert = k-1$:} Let there be $X \subset \mathcal{S}$ with $\lvert X \rvert = k-1$ such that HC2($X$) is violated. Let us assume that $r$ sets of $\mathcal{X}$ are contained in $X$, where $r \geq k$. Let among those $r$ sets $U_1, U_2, \ldots, U_q$ are $(k-2)$-sets, and $V_1, V_2, \ldots, V_{r-q}$ are $(k-3)$-sets, for some $q \leq r$.\par
For each $V_i$, $1 \leq i \leq r-q$, let $\mathcal{H}_i = \{W \mid V_i \subset W \subset X\}$. That is, $\mathcal{H}_i$ is a set of $(k-2)$-subsets of $X$ that contain $V_i$. Now, observe the following.
\begin{enumerate}
\item $\lvert \mathcal{H}_i \rvert =2, 1\leq i \leq r-q$.
\item $U_i \notin \mathcal{H}_j, 1 \leq i \leq q, 1 \leq j \leq r-q$. This is because, following construction, $V_j \nsubseteq U_i, 1 \leq i \leq q, 1 \leq j \leq r-q$.
\item $\mathcal{H}_i \cap \mathcal{H}_j = \emptyset$ if $V_i \neq V_j, 1 \leq i < j \leq r-q$. As noted earlier in construction step 1, this follows from our choice of auxiliary collection $\mathcal{X}_a$; no two distinct $(k-3)$-sets of $\mathcal{X}$ have the same $(k-2)$-superset.
\end{enumerate}
Since there are $2$ copies of each $(k-3)$-subset in the final collection $\mathcal{X}$, there are at least $\lceil \frac{r-q}{2} \rceil$ distinct $V_i$ s in $\mathcal{X}$. Hence, there are at least $\lceil \frac{r-q}{2} \rceil$ disjoint $\mathcal{H}_i$ s. Thus, number of distinct $(k-2)$-subsets of $X$ is at least $q + 2 \lceil \frac{r-q}{2} \rceil \geq r \geq k$, which is not possible since $\lvert X \rvert = k-1$. \\
Thus $X$ satisfies HC2($X$). \qed
\end{enumerate}
\par
So, $(\mathcal{S}, \mathcal{X})$ is an $(n, N, k, m)$-CBC, where $N = \sum_{X \in \mathcal{X}} \lvert X \rvert = n(k-2) - 2 \left\lfloor \frac{U_{m, k, k-2} - n}{m-k+1} \right\rfloor$. \par
Regarding optimality of the construction, we note that lower bound on $N(n, k, m)$, given by Theorem \ref{th2} for this range of values of $n$ is $n(k-2) - \left\lfloor\frac{2(U_{m, k,k-2}-n)}{m-k+1}\right\rfloor$. Hence, difference between value of $N$ obtained from our construction and value of $N$ given by the lower bound that we have obtained, for given values of $n, m,$ and $k$ is\\
\vspace{5pt}
$\left(n(k-2) - 2 \left\lfloor \frac{U_{m, k, k-2}-n}{m-k+1} \right\rfloor\right)- \left(n(k-2) - \left\lfloor\frac{2(U_{m, k, k-2} - n)}{m-k+1}\right\rfloor\right)\\
= \left\{
\begin{array}{l l}
0 \quad \text{when} \quad0 \leq (U_{m, k, k-2} - n) \mod (m-k+1) < \frac{m-k+1}{2},\\
1 \quad \text{when} \quad \frac{m-k+1}{2} \leq (U_{m, k, k-2} - n) \mod (m-k+1) < m-k+1.\\
\end{array}
\right.$
\\
So, the construction yields optimal CBCs for approximately half of the values of $n$ in this range; for the other half, value of $N$ for constructed CBC differs by one from the corresponding lower bound that we have obtained. Hence, we have the following theorem.
\begin{theorem}
Let $\binom{m}{k-2}-(m-k+1)A(m, 4, k-3) \leq n \leq \binom{m}{k-2}$. Then $N(n, k, m) = n(k-2) - \left\lfloor\frac{2(\binom{m}{k-2} - n)}{m-k+1}\right\rfloor$ for $0 \leq (\binom{m}{k-2} - n)\mod (m-k+1) < \frac{m-k+1}{2}$ and $N(n, k, m) \leq n(k-2) - 2\left\lfloor\frac{\binom{m}{k-2} - n}{m-k+1}\right\rfloor$ for $\frac{m-k+1}{2}\leq (\binom{m}{k-2} - n)\mod (m-k+1) < m-k+1$.
\end{theorem}
\subsection{Construction of $c$-uniform CBCs for $1 \leq \lfloor \frac{k}{2}\rfloor \leq c < k-1$}
In this section, we give explicit construction of a $c$-uniform $(n, cn, k, m)$-CBC $(\mathcal{S}, \mathcal{X})$ for given values of $k$ and $m$, where $1 \leq \lfloor \frac{k}{2}\rfloor \leq c < k-1$ and $n= (k-c-1) A(m, 2(k-c-1), c)$.\par
\cn Let $\mathcal{S}$ be the set of servers, where $\lvert \mathcal{S} \rvert = m$.  $\mathcal{X}$ is a collection of $c$-subsets of $\mathcal{S}$, each having $k-c-1$ copies, where distinct $c$-subsets correspond to codewords of a binary constant weight code of length $m$, weight $c$, and distance $2(k-c-1)$. So, distinct $c$-sets of $\mathcal{X}$ are mutually minimum $2(k-c-1)$ distance apart. Note that for $1 \leq \lfloor \frac{k}{2}\rfloor \leq c < k-1$, we have $1 \leq k-c-1 \leq \lfloor \frac{k-1}{2} \rfloor \leq \lfloor \frac{k}{2}\rfloor \leq c $. \par 
This construction yields $c$-uniform $(n, cn, k, m)$-CBC $(\mathcal{S}, \mathcal{X})$, where $n = \lvert \mathcal{X} \rvert = (k-c-1) A(m, 2(k-c-1), c)$. Below we prove correctness of the construction.\par 
\pocr We prove correctness of this construction by showing that $(\mathcal{S}, \mathcal{X})$ satisfies HC2[$k$], i.e., HC2($X$) is satisfied for all $X \subset \mathcal{S}$ with $0\leq \lvert X \rvert \leq k-1$.\par 
\begin{enumerate}
\item {\bf HC2($X$) for $X \subset \mathcal{S}$ with $0 \leq \lvert X \rvert \leq k-2$:} HC2($X$) is trivially satisfied for all $X \subset \mathcal{S}$ with $0 \leq \lvert X \rvert \leq c-1$. Since there are $k-c-1$ copies of each $c$-set and $c \geq k-c-1$, HC2($X$) is also satisfied for all $X \subset \mathcal{S}$ with $\lvert X \rvert = c$. Now, union of two distinct $c$-sets of $\mathcal{X}$ contains exactly $k-1$ elements. Hence HC2($X$) is satisfied for all $X \subset \mathcal{S}$ with $c+1 \leq \lvert X \rvert \leq k-2$.
\item {\bf HC2($X$) for $X \subset \mathcal{S}$ with $\lvert X \rvert = k-1$:} For any $X \subset \mathcal{S}$ such that $\lvert X \rvert = k-1$, one of the following applies.
\begin{enumerate}
\item $X$ contains at most two distinct $c$-sets of $\mathcal{X}$. In this case, $X$ contains at most $2(k-c-1) \leq 2 \lfloor \frac{k-1}{2} \rfloor \leq k-1$ $c$-sets of $\mathcal{X}$. Hence HC2($X$) is satisfied for this case.
\item $X$ contains at least $3$ distinct $c$ sets of $\mathcal{X}$. Then we have
\begin{enumerate}
\item for any two distinct $X_1, X_2 \in \mathcal{X}$ such that $X_1, X_2 \subseteq X$, it follows that $\lvert X_1 \cup X_2 \rvert= \lvert X \rvert = k-1$ and $\lvert X_1 \setminus X_2\rvert = k-c-1$;
\item  for any three distinct $X_1, X_2, X_3 \in \mathcal{X}$ such that $X_1, X_2, X_3 \subseteq X$, $(X_1 \setminus X_2) \cap (X_1 \setminus X_3) = \emptyset$. For, if for some $s \in X$, $s \in (X_1 \setminus X_2) \cap (X_1 \setminus X_3)$, then $s \notin (X_2 \cup X_3)$. Hence, $\lvert X_2 \cup X_3 \rvert \leq k-2$, a contradiction.
\end{enumerate}
Hence, given $Y \in \mathcal{X}$ such that $Y \subseteq X$, there can be at most $\lfloor \frac{c}{k-c-1} \rfloor$ other distinct $X'$ such that $X' \in \mathcal{X}$, $X' \subseteq X$, and $\lvert X' \setminus Y \rvert = k-c-1$. Hence, there are at most $(k-c-1)\lfloor \frac{c}{k-c-1}\rfloor+k-c-1 \leq k-1$ $c$-sets of $\mathcal{X}$ contained in $X$. Hence, HC2($X$) is satisfied for this case.
\end{enumerate}
So, HC2($X$) is satisfied for all $X \subset \mathcal{S}$ with $\lvert X \rvert = k-1$. \qed
\end{enumerate}
In order to get an idea of the value of $n$, i.e., size of $\mathcal{X}$, we use asymptotic bound given in Theorem \ref{approx}, which shows $n \gtrsim \frac{(k-c-1)m^{2c-k+2}}{c!}$ as $m \to \infty$. In \cite{PaStWe}, the authors gave non-constructive proof of existence of $c$-uniform $(n, cn, k, m)$-CBCs, for which $n$ is $\Omega(m^{\frac{ck}{k-1}-1})$; for sufficiently large $m, c,$ and $k$ such that $c \sim k$ and $c^c \sim m$, value of $n$ (in asymptotic sense) obtained from our explicit construction compares well with this value.
%\noindent 

%{\bf Acknowledgements:} We wish to thank the reviewers 
%for reading the paper and providing several suggestions.

\end{document}